# CONFINING POTENTIAL AND MASS OF ELEMENTARY PARTICLES

L.I. Buravov

Institute of Problems of Chemical Physics, RAS

buravov@icp.ac.ru

*Abstract*

In this paper we consider a model in which the masses of elementary particles are formed and stabilized thanks to confining potential, which is caused by recoil momentum at emission of specific virtual bosons by particle itself. The calculation of this confining potential $\Phi(R)$ is carried out. It is shown that $\Phi(R)$ may be in the form const $R^3$ or const $R^2$ depending on continuous or discrete nature of the spectrum of emitted bosons.

In the article the model of formation of mass of elementary particles is offered as a result of emitting of the special virtual bosons as spherical waves (conditionally we will name them as the bosons of Higgs). It is shown in the article, as confining potential necessary for stabilizing of particle mass appears because of effect of impulse recoil, in particular for electron, muon, pion, kaon and neutrino.

In due time Poincare, proceeding from common sense, entered supposition about a presence in the structure of electron of some elastic element due to which the charge of electron holds out in a small volume. This model was later used by many authors. We will consider it more detailed, following [1,2].

In accordance with [1,2] virtual rest energy of electron $E$ consists of two parts: surface energy of elastic shell $W_s = \sigma 4\pi R^2$ and electrostatic energy of the charged surface $W_e = e^2/2R$, where $\sigma$-is the coefficient of surface tension of the shell, $R$ is its radius, $e^2 = q^2/4\pi\varepsilon_o$, $\varepsilon_o$ is dielectric constant of vacuum, $q$ is an electron charge (in units of SI). Virtual rest energy of such system is equal:

$$E = \sigma 4\pi R^2 + e^2/2R \qquad (1)$$

The radius of electron, corresponding to a minimum of energy (1) of the system, is determined from equation $\partial E/\partial R = 0$ and is equal:

$$R_e = 0.5(e^2/2\pi\sigma)^{1/3} \qquad (2)$$

From equations (1) and (2) the mass of electron is

$$m_e = 3(\pi\sigma/4)^{1/3} e^{4/3}/c^2 \qquad (3)$$

It is possible also to write down that $m_e = 12\sigma\pi R_e^2/c^2 = f R_e^2$, where a $\sigma$ value was determined in [3] with using the value of neutral pion mass $m_o = 134.963$ Mev/$c^2$ :

$$\sigma = 4 \cdot 3^{-7}\pi^{-3} \hbar^{-2}c^4 m_o^3 = 5.967 \cdot 10^{14} \text{ J/m}^2 \qquad (4),$$

and coefficient $f = 12\sigma\pi/c^2$ ( $\cong 1.404 \cdot 10^{25}$ Mev/cm$^2$ ).

In equation (1) the value $W_S = \sigma 4\pi R^2$ obviously fulfills the role of confining potential due to which the mass of electron is stabilized.

Before to pass to more detailed consideration of confining potential, preliminary we will represent information about the calculation of the masses for several elementary particles.

In [3] the calculation of the ratio for the masses of the particles e, $\mu$, $\pi^0$, $\pi^{\pm}$, K$^0$, K$^{\pm}$ was made on the basis of starting model assumption, that the stopped muon, pion and kaon can be represented as spherical resonators for quanta of virtual neutrinos excited into an elastic lepton shell with surface energy $W_S = \sigma 4\pi R^2$, where $R$ is a radius of elastic shell, $\sigma$-the coefficient of surface tension, the same, that in equations (1-4). In [3] it was shown that virtual rest energy of these particles in general case can be written down in some more complex form, than for an electron :

$$E = \sigma 4\pi R^2 + 1.5N\hbar\pi c/(R-\rho) + e^2/2\rho \qquad (5),$$

where $\rho$ is a radius of the compressed electric cloud, and $N$ is a number of neutrino quanta that is determined from the decay scheme: $N = 2$ for muon, 3-for pion, and 21 - for kaon. The masses of particles and characteristic sizes $R_m$ and $\rho_m$ are determined in general case at minimization of virtual rest-energy (5) on $R$ and $\rho$. Calculated in [3] values of the masses of e, $\mu$, $\pi^0$ and K$^0$ are in relation 0.547: 105.707: 134.963: 493.87 (Mev) (by attachment to mass of neutral pion), that is in accord with experience data. It was shown that the masses of all considered particles, as well as for an electron, are proportional to the square of their equilibrium size $R_m$ :

$$Mc^2 = f c^2 R_m^2 = 12\sigma\pi R_m^2 \qquad (6)$$

It is assumed in the Standard Model, that the masses of row of elementary particles can be represented in the form :

$$Mc^2 = \lambda_M H/\sqrt{2} \qquad (7)$$

where $H=246$ Gev is characteristic energy in the model of Higgs [4,5], $\lambda_M$ is a dimensionless factor, characteristic for a certain particle with mass of $M$. One can formula (6) for the masses, got in [3] make compatible with (7), if preliminary to equate right parts (6) and (7); that results to:

$$\lambda_M = R_m^2 /[ H/(\sqrt{2} \cdot 12\sigma\pi)] = R_m^2/R_x^2 \qquad (8),$$

where unknown size of $R_x=[H/(\sqrt{2} \cdot 12\sigma\pi)]^{1/2}=1.11 \cdot 10^{-10}$ cm , that is comparable to ½ the Compton wave-length of electron $2\pi\hbar/m_e c = 2.43 \cdot 10^{-10}$ cm.

The result of the formula (6) was used in [6] for the calculation of neutrino masses $\nu_e$, $\nu_\mu$ and $\nu_\tau$, for which the square of their electromagnetic radius was found in works [7-9]. It was shown in them, that neutrinos of all types have a complex internal structure as a consequence of the virtual transitions $\nu_\ell \leftrightarrow \ell^- + W^+$, $\tilde{\nu}_\ell \leftrightarrow \ell^+ + W^-$, where the lepton index $\ell$ means e, μ or τ ; W are intermediate vector bosons, carriers of the weak interaction with mass of $M_w = 80.4$ Gev [10]. Taking into account such virtual transitions, in [7-9] it was found that the square of the electromagnetic radius of neutrino is equal to:

$$<r^2(\nu_\ell)>=(3G_F/8\sqrt{2}\,\pi^2\hbar c)[(5/3)\mathrm{Ln}\alpha + (8/3)\mathrm{Ln}(M_w/m_l) +\eta] \qquad (9),$$

where $G_F = 1.43 \cdot 10^{-62}$ J·m$^3$ is the weak interaction constant, $\alpha = e^2/\hbar c \cong 1/137$, and numerical constant $\eta$ is 1-2. For a mean value $\eta = 1.5$ taking into account $m_\mu c^2 =105.66$ Mev and $m_\tau c^2 =1777$ Mev from (9) it follows that the characteristic values of the squared neutrino radii are equal to:

$$<r^2(\nu_e)> \cong 3 \cdot 10^{-33} \mathrm{cm}^2, <r^2(\nu_\mu)> \cong 1.3 \cdot 10^{-33}\mathrm{cm}^2, <r^2(\nu_\tau)> \cong 4.2 \cdot 10^{-34}\mathrm{cm}^2 \qquad (10)$$

To define the masses of neutrino, in [6] simple suppositions were made:

1.Although neutrinos do not have an electric charge, but they seems to have small electrostatic energy due to that spacial distributions of diverse charges produced by virtual pairs ($\ell$, W) are slightly different. In this case the electrostatic energy of neutrino has a value $U(\nu_\ell) = \delta(\nu_\ell)e^2/r$, where $r$ is a virtual electromagnetic radius of neutrino, $\delta(\nu_\ell)$ is an unknown small parameter related to the distribution of charges in a structure of $\nu_\ell$.

2.Virtual rest-energy of neutrino consists of the confining potential $W_S= \sigma 4\pi r^2$ and electrostatic energy :

$$E = \sigma 4\pi r^2 + \delta(\nu_\ell)e^2/r \qquad (11)$$

3. The value of σ is identical for all neutrinos.

Similarly, as for an electron, mass of neutrino would be found at being of a minimum of virtual energy (11) :

$$m(\nu_\ell)=3(\pi\sigma)^{1/3}[\delta(\nu_\ell)e^2]^{2/3}/c^2 \qquad (12),$$

but as a value $\delta(\nu_\ell)$ is unknown, for determination of the neutrino masses we will take advantage of theoretical results [7-9] for the square of electromagnetic radius of neutrino $<r^2(\nu_\ell)>$ and of a formula (6).Thus, putting the found values (10) for $<r^2(\nu_\ell)>$ in a formula (6) instead of $R_m^2$, we find that the masses of neutrino are equal:

$$m(\nu_e)c^2 \cong 4.3 \cdot 10^{-2} \mathrm{eV}, \quad m(\nu_\mu)c^2 \cong 2\cdot 10^{-2} \mathrm{eV}, \quad m(\nu_\tau)c^2 \cong 6\cdot 10^{-3} \mathrm{eV} \qquad (13)$$

Similar values for three neutrino mass eigenstates ($\nu_1,\nu_2,\nu_3$) were received in [11] on the basis of Super-Kamiokande experimental results [12], inventively solving system of two equations with 3 unknown quantities, if supposing the case of inverted mass spectrum.

Values $\lambda(\nu_\ell)$ for a neutrino also can be found from a formula (8) at the substitution of values $<r^2(\nu_\ell)>$ instead of $R_m^2$:

$$\lambda(\nu_e) \cong 2.47 \cdot 10^{-13}, \lambda(\nu_\mu) \cong 1.08 \cdot 10^{-13}, \quad \lambda(\nu_\tau) \cong 3.4 \cdot 10^{-14} \qquad (14)$$

We will notice that as $G_F \cong \pi\alpha(\hbar c)^3/(\sqrt{2}\sin^2\theta_w M_w^2 c^4)$ [5,13], taking into account equations (4), (6) and (9) we get a general formula for the masses of neutrino in the form:

$$m(\nu_\ell) \cong H(\ell)\alpha[m_o/M_w]^2 m_o \qquad (15),$$

where a dimensionless factor $H(\ell)$ is equal:

$$H(\ell)=3^{-5}\pi^{-3}\sin^{-2}\theta_w[(5/3)\mathrm{Ln}\alpha +(8/3)\mathrm{Ln}(M_w/m_\ell) +\eta] \qquad (16)$$

$\theta_w$ –is the angle of Weinberg ($\sin^2\theta_w= 0.23$ ), and all 3 dimensionless coefficients: $H(\ell),\alpha$ and $[m_o/M_w]^2$ are much smaller than 1.

The values $\delta(\nu_\ell)$ are determined from the equation (12):

$$\delta(\nu_\ell) = [m(\nu_\ell)c^2]^{3/2}/[3^{3/2}\alpha\hbar c(\pi\sigma)^{1/2}] \qquad (17)$$

and are equal:

$$\delta(\nu_e) \cong 1.10 \cdot 10^{-11}, \delta(\nu_\mu) \cong 3.17 \cdot 10^{-12}, \delta(\nu_\tau) \cong 5.6 \cdot 10^{-13} \qquad (18)$$

In equation (5), also as in (1), the value $W_S = \sigma 4\pi R^2$ plays the role of a confining potential thanks to which the mass of the elementary particle is stabilized. This conception of the mass origin is simple and clear: complete internal energy is equal $Mc^2$; however, physical reason for the origin of confining potential is not clear. A model is offered in this article, explaining the origin of this potential and holding pressure due to the effect of impulse recoil of the special emitted virtual bosons.

So then for the ground of this model we will enter next suppositions:

1. By analogy with the model of Poincare we will suppose that every elementary particle has confining potential $\Phi(R)$, except for photons and gravitons. It is assumed that rest-energy of particle consists of confining potential, kinetic energy of internal motion, energy of the internal fields and electrostatic energy of the internal charges.

2. Every elementary particle radiates the special virtual bosons as spherical waves $A(e^{ikr}/r)e^{-i\omega t}$. Here $k$ - is a wave-number of virtual boson, $\omega$ – its angular frequency, $r$ is distance from the center of a particle ($r \geq R$), $A$ is a normalizing constant not substantial for further consideration.

3. We will suppose that mass of such bosons $M_H$ is much more than masses of intermediate vector bosons W and $Z^0$, where $Z^0$ is a neutral carrier of the weak interaction with mass 91.2 Gev [10]. Then a time of existence of such virtual bosons $\tau$ is much smaller than $2\pi\hbar/(M_w c^2) \approx 5 \cdot 10^{-26}$ sec and distance of their run of $L_\tau$ from the surface of elementary particle is much smaller than $2\pi\hbar/(M_w c) \approx 2 \cdot 10^{-15}$ cm.

4. Every elementary particle is the inehaustible source of such virtual spherical waves, but the mass of the particle-source does not decrease, because virtual bosons through an instant $\tau$ return back into a source, due to interacting with the fields of vacuum.

5. We suppose that complete amount of the bosons emitted by a particle in a unit of time $N_H$ is proportional to area of particle surface with a coefficient $\gamma$, characterizing intensity of radiation for the certain group of particles: $N_H = \gamma 4\pi R^2$.

As every moving wave carries an impulse, it ensues from these suppositions, that on the surface of elementary particle because of the effect of impulse recoil, spherical waves are creating holding force of pressure $F(R)$ and confining potential $\Phi(R) = \int F(R)dR$.

We will consider 2 cases for forms of boson spectrum:

a) the emitted bosons have a continuous spectrum of radiation in the interval of $k$ from 0 to $k_{max} = K$ with the function of probability distribution $W_a(k, \xi)$, where $\xi = M_H c/\hbar$. As be shown, in this case confining potential is proportional to $R^3$.

b) the emitted bosons have a discrete spectrum of radiation : $k = \pi/R, 2\pi/R,...n\pi/R$ ...with the function of probability distribution $W_b(k,\xi)$. In this case the confining potential appears to be proportional to $R^2$.

Let's consider the case of a). We will enter the condition of normalizing for $W_a(k, \xi)$, so that $\int_0^K W_a(k,\xi)dk = 1$. Then the number of impulses in an interval from $k$ to $k+dk$ is equal $N_H W_a(k, \xi)dk = \gamma 4\pi R^2 W_a(k, \xi)dk$, and complete force of pressure, operating on a surface $4\pi R^2$ is equal :

$$F(R) = 2\int_0^K \hbar k \gamma 4\pi R^2 W_a(k,\xi)dk \qquad (19)$$

Consequently confining potential for the case of a) is :

$$\Phi_3(R) = R^3 [(8/3)\pi\gamma\hbar \int_0^K k\, W_a(k,\xi)dk] = R^3[(8/3)\pi\gamma\hbar k_{av}] \qquad (20),$$

where $k_{av} = \int_0^K k\, W_a(k,\xi)dk$ is a quantum-mechanical mean value of $k$.

For the case of b) we will enter the condition of normalizing for $W_b(k, \xi)$ : $\sum_n W_b(n\pi/R, \xi) = 1$. In this case total force of pressure on a surface $4\pi R^2$ is equal:

$$F(R) = 2\sum_n \hbar(n\pi/R)\gamma 4\pi R^2 W_b(n\pi/R,\xi) \qquad (21).$$

The sum $\sum_n (n\pi/R) W_b(n\pi/R,\xi)$ in the formula (21) by definition is equal to $n_{av}(\pi/R)$, where $n_{av}$ is a quantum-mechanical mean value of $n$. As a result the confining potential $\Phi_2(R)$ is :

$$\Phi_2(R) = R^2 [4\pi^2 \gamma\hbar n_{av}] \qquad (22).$$

Thus, in the article assumed, that every elementary particle produces the special bosonic field that is present only in a thin layer at the surface of a particle. It is shown, that this field can create the confining potential, stabilizing the mass of particle during the time of its life.